\documentclass[aps,prl,twocolumn,superscriptaddress,floatfix]{revtex4-1}

\usepackage{amsmath}
\usepackage{amssymb}
\usepackage{amsfonts}
\usepackage{listings}
\usepackage{enumerate}
\usepackage{latexsym}
\usepackage{bm}
\usepackage{graphicx}
\usepackage{braket}
\usepackage[pdftex,colorlinks=false]{hyperref}
\usepackage{xcolor}
\usepackage{verbatim}
\usepackage{float}

\usepackage{times}
\usepackage[normalem]{ulem}   
\usepackage{cancel,soul}
\usepackage{manfnt}   
\usepackage{txfonts}  
\usepackage{amsbsy}

\newcommand{\beq}{\begin{equation}}
\newcommand{\eneq}{\end{equation}}

\def\:={\,\raisebox{0.95pt}{.}\hspace{-2.48pt}\raisebox{2.85pt}{.}\!\!=\,}
\def\=:{\,=\!\!\raisebox{0.85pt}{.}\hspace{-2.78pt}\raisebox{2.85pt}{.}\,}

\newcommand*{\boldplus}{\ensuremath{\boldsymbol{\pmb{+}}}}
\newcommand*\cube{\mbox{\mancube}}

\begin{document}

\tolerance 10000

\newcommand{\cbl}[1]{\color{blue} #1 \color{black}}

\newcommand{\vk}{{\bf k}}

\widowpenalty10000
\clubpenalty10000

\title{
Topological many-body scar states in dimensions 1, 2, and 3
}

\author{
Seulgi~Ok}
\affiliation{
 Department of Physics, University of Zurich, Winterthurerstrasse 190, 8057 Zurich, Switzerland
}

\author{
Kenny~Choo}
\affiliation{
 Department of Physics, University of Zurich, Winterthurerstrasse 190, 8057 Zurich, Switzerland
}

\author{
Christopher Mudry}
\affiliation{
Condensed Matter Theory Group, Paul Scherrer Institute, CH-5232 Villigen PSI, Switzerland}
\affiliation{
Institute of Physics,
\'Ecole Polytechnique F\'ed\'eerale de Lausanne (EPFL), CH-1015 Lausanne, Switzerland
}

\author{
Claudio Castelnovo}
\affiliation{
Theory of Condensed Matter Group, Cavendish Laboratory,
University of Cambridge, Cambridge CB3 0HE, United Kingdom}

\author{
Claudio Chamon}
\affiliation{
Physics Department, Boston University, Boston, Massachusetts 02215, USA}

\author{
Titus~Neupert}
\affiliation{
 Department of Physics, University of Zurich, Winterthurerstrasse 190, 8057 Zurich, Switzerland
}

\begin{abstract}
We propose an exact construction for atypical excited states of a
class of non-integrable quantum many-body Hamiltonians in
one dimension (1D), two dimensions (2D), and three dimensins (3D)
that display area law entanglement entropy. These examples of
many-body ``scar'' states have, by design,
other properties, such as topological degeneracies,
usually associated with the gapped ground states of
symmetry protected topological phases or topologically ordered phases
of matter.
\end{abstract}

\date{\today}

\maketitle

\textit{Introduction} ---
Until recently, the study of many-body quantum systems has largely focused on
ground-state properties and low-energy excitations, implicitly assuming
the eigenstate thermalization hypothesis (ETH) dictating that highly
excited states of generic non-integrable models are void of interesting
structures~\cite{berry1977level,SrednickiETH1994}.
With the discovery of quantum systems that violate the ETH,
a broader interest in the physics of many-body excited states emerged. 
This modern development is complemented by the
growing potential of quantum simulators -- predominantly using
ultracold atomic gases -- to prepare and study
quantum many-body systems that are well isolated from the
environment~\cite{kinoshita2006quantum,schreiber2015observation}.

Theoretical indicators for the violation of the ETH by
a conserved quantum
many-body Hamiltonian include
(i) a sub-volume law scaling for the entanglement entropy of eigenstates,
(ii) emergent local integrals of motion in a non-integrable system~\cite{geraedts2017emergent,imbrie2017local},
and (iii) oscillations in the expectation value
of suitably chosen local observables under the unitary time-evolution~\cite{turner2018weak}.

Two examples of ETH-violating conserved quantum Hamiltonians
are those that either support
(1) quantum many-body localized states%
~\cite{fleishman1980interactions,gornyi2005interacting,
basko2006metal,oganesyan2007localization,pal2010many,
AbaninMBL2013,HuseMBL2014,imbrie2016many},
where nearly all eigenstates at finite energy
density share properties (i) and (ii),
and (2) many-body quantum scars,
where only a small set of states embedded in a
continuum of thermalizing states show such exotic behavior%
~\cite{vafek2017entanglement,shiraishi2017systematic,ho2018periodic,
turner2018weak,moudgalya2017exact,moudgalya2018entanglement,
turner2018quantum,lin2018exact}.
Here, we will be concerned with examples for the latter.

Theoretical studies of such ETH-violating systems are challenging for two
reasons. Analytical progress%
~\cite{basko2006metal,imbrie2016many,vafek2017entanglement,
moudgalya2017exact,moudgalya2018entanglement,lin2018exact}
is hard because the models in question
are, by definition, non-integrable.
Numerical techniques to obtain highly excited states
rely on exact-diagonalization~\cite{ShiftInverse}
and, in some cases, matrix-product state calculations%
~\cite{DMRGX}.
These techniques are limited in that the range of
available system sizes is often too small to allow an extrapolation
to the thermodynamic limit.  For these reasons, the majority of
studies on ETH-violation have been focused on one-dimensional (1D) models.

In this work we present a generic construction that places a scar
state in the spectrum of non-integrable many-body quantum systems in
1D, 2D, and 3D. While the construction of such states
applies to many systems,
our primary focus is on topological scar states.
In 1D, we construct symmetry-protected topological (SPT)
states~\cite{Chen2013}. In 2D, we present a non-integrable deformation
of the toric-code, with 4-fold degenerate scar states on the
torus. Finally, in 3D we present a deformation of the X-cube
model~\cite{Castelnovo2010,SagarXCube2016} as an example of a system
with scars that display fracton topological
order~\cite{Chamon2005,Bravyi2011,Haah2011,Castelnovo2010,SagarFracton2015,SagarXCube2016}.

Our construction is inspired by families of Hamiltonians that have
been studied in the contexts of quantum dimer models and spin
liquids~\cite{RokhsarKivelson1988,Sachdev1989,IoffeLarkin1989,Henley1997,MoessnerSondhi2001,Castelnovo2005,ChamonCastelnovo2008}. In those studies, the emphasis was on the construction of parent
Hamiltonians for a given ground state. Consider the Hamiltonian
\begin{subequations}
\begin{align}
H(\beta)\:=
\sum_{s}
\alpha^{\,}_{s}\, Q^{\,}_{s}(\beta),
\label{eq:1a Hamiltonian}
\end{align}
where $s$ labels certain bounded regions of space,
such as the elementary plaquettes of a lattice. The operators
$Q^{\,}_{s}(\beta)$ are Hermitian, positive-semidefinite, and local
(i.e., with bounded and discrete spectra), and contain
only sums of products of operators defined within the bounded region
labeled by $s$. A family of such local operators is parametrized by
the dimensionless number $\beta$,
which we shall later deploy to deform solvable models and
break integrability. The dimensionfull coupling constants
$\alpha^{\,}_{s}\in\mathbb{R}$ carry the units of energy.
The operators $Q^{\,}_{s}(\beta)$
are built so as to share a common null state
$|\Psi(\beta)\rangle$, i.e.,
\begin{align}
Q^{\,}_{s}(\beta)\;|\Psi(\beta)\rangle=0,
\quad \forall\;s.
\label{eq:1b Hamiltonian}
\end{align}
\end{subequations}
[For instance, at the Rokhsar-Kivelson point of the quantum dimer
 model on the square lattice,
$s$ would be a plaquette and the operators $Q^{\,}_{s}(\beta)$ are
projectors that encode both the potential and kinetic (plaquette flip)
terms~\cite{RokhsarKivelson1988,Henley1997,Castelnovo2005}.]
If all the couplings $\alpha^{\,}_{s}$ are
positive, the state $|\Psi(\beta)\rangle$ is the ground state of
$H(\beta)$, as the $Q^{\,}_{s}(\beta)$ are positive-semidefinite. If the
$\alpha^{\,}_{s}$ take positive or negative values depending on $s$, then
one cannot guarantee that $|\Psi(\beta)\rangle$ is a ground state. It
is, nonetheless, an eigenstate with energy $E=0$. Even when this state
is a high energy eigenstate in the spectrum of $H(\beta)$, it is an
atypical state in that it displays area law entanglement entropy, for
it is also a ground state of a \emph{different} local Hamiltonian
$\widehat{H}(\beta)\:=\sum_{s} |\alpha^{\,}_{s}|\, Q^{\,}_{s}(\beta)$. Hence
$|\Psi(\beta)\rangle$ is a scar state, if $H(\beta)$ is nonintegrable. 
(Reference~\onlinecite{shiraishi2017systematic} also presents an analytical 
construction of scar states; we explain the connection in the Supplemental 
Information.) 

By deforming exactly solvable models -- the toric code, for instance
-- one can break integrability while retaining the $E=0$ scar
state. [In the Supplementary Material we show how to construct
non-commuting $Q^{\,}_{s}(\beta)$ operators with the desired properties
starting from solvable models with commuting projectors.] In what
follows, we construct topological scar states in 1D, 2D, and 3D.

\textit{A warm-up example ---} We start with a simple example in 1D,
which is topologically trivial, but illustrates the general ideas in a
straighforward way. Consider a quantum spin-1/2 1D chain with periodic
boundary conditions, i.e., a ring, with $L$ sites. On each site $i =
1,\cdots, L$, we denote the three Pauli operators by $X^{\,}_{i}$,
$Y^{\,}_{i}$, and $Z^{\,}_{i}$.
For any $\beta\ge0$, we define the local Hamiltonian
\begin{subequations}
\label{eq:1D Hamiltonian}
\begin{align}
&
H(\beta)\:=
\sum_{i}
\alpha^{\,}_{i}\,
Q^{\,}_{i}(\beta),
\label{eq:1D Hamiltonian a}
\\
&
\alpha^{\,}_{i}\:=
\alpha+(-1)^{i},
\qquad
Q^{\,}_{i} (\beta)\:=
e^{-\beta\,(Z^{\,}_{i-1}\,Z^{\,}_{i}+Z^{\,}_{i}\,Z^{\,}_{i+1})}
-
X^{\,}_{i},
\label{eq:1D Hamiltonian b}
\end{align}
\end{subequations}
with $0<|\alpha|<1$. The condition $|\alpha|<1$ is required to place the scar
state in the middle of the spectrum; the condition $\alpha \ne 0$ is needed so
as not to break the system into two independent (and integrable)
transverse-field Ising chains.

\begin{figure}[t]
\centering
\includegraphics[width=0.46\textwidth]{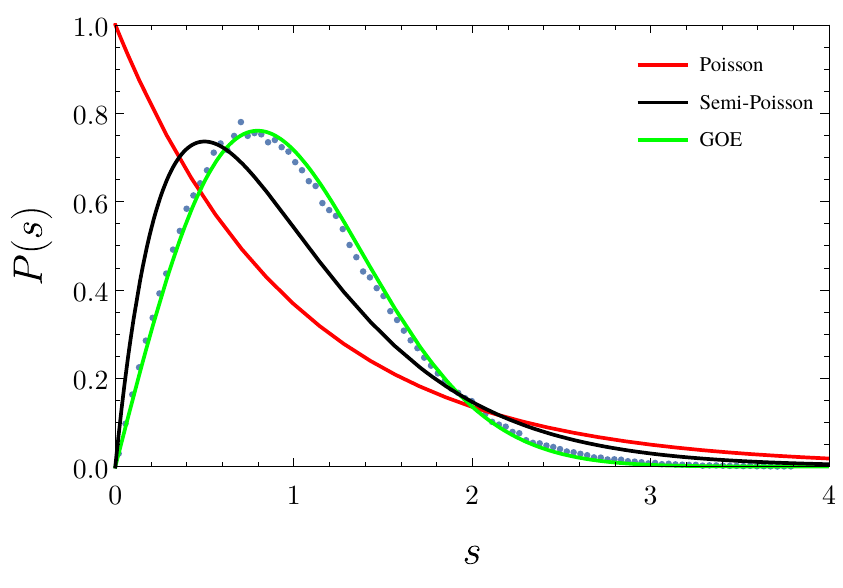}
\caption{
(Color online)
Distribution of consecutive energy level spacings $s^{\,}_{n}$ for
the 1D Hamiltonian $H$ defined in Eq.~\eqref{eq:1D Hamiltonian} with
$L=20$, $\alpha=0.3$, and $\beta=0.5$. 
The distributions for the $s^{\,}_{n}$ from all momentum sectors, except for 
$k=0,\pi$, have been joined. 
The middle 60\% of the spectrum in each sector is taken.
The distribution obtained can be seen to be well approximated by the 
Gaussian orthogonal ensemble (GOE) of random matrix theory. 
        }
\label{fig:level statistics}
\end{figure}

At $\beta=0$, the system is equivalent to a paramagnetic spin chain in
a Zeeman field, which is integrable.
With $\beta\ne0$, all the nearest-neighbor terms no longer commute, i.e.,
$[Q^{\,}_{i}(\beta),Q^{\,}_{i\pm1}(\beta)]\ne0$.
In this case, $H(\beta)$ should no longer be integrable,
a fact confirmed by analysis of the energy level statistics obtained
numerically as we now explain.
We study the statistics of the spacings between
consecutive energy levels, $s^{\,}_{n}\:= E^{\,}_{n+1}-E^{\,}_{n}$, as well as the
$r$-value defined as the average 
$\langle r^{\,}_{n}\rangle$
of the ratios 
$r^{\,}_{n}\:={\min(s^{\,}_{n},s^{\,}_{n-1})}/{\max(s^{\,}_{n},s^{\,}_{n-1})}$. 
We analyze the spectrum in common eigenspaces of a maximal set of commuting
symmetries of the system, namely
translation, parity under inversion, and an additional
$\mathbb{Z}_{2}$-valued parity defined by
$\prod_{i}X^{\,}_{i}=\pm1$.
Figure~\ref{fig:level statistics} contains the result of this analysis
for $\alpha=0.3$, $\beta=0.5$ and $L = 20$. The distribution matches
the distribution of eigenvalue spacings for the Gaussian Orthogonal
Ensemble (GOE) of random matrices, thus supporting the claim that
Hamiltonian (\ref{eq:1D Hamiltonian}) is non-integrable.
The corresponding mean $r$-value for our distribution (averaged over the different momentum sectors) is $\langle r \rangle = 0.531$, close to that of the GOE, $r_{\mathrm{GOE}} = 0.5359$,
and clearly distinct from the value of the Poisson distribution, $r_{\textrm{Poisson}}=0.3863$.

One can verify that the state
\begin{subequations}
\label{eq:1Dscar3a3b}
\begin{align}
|\textrm{scar}(\beta)\rangle\:=
G(\beta)\;
\bigotimes_{i}\;
|+\rangle^{x}_{i},
\label{eq:analytical scar general}
\end{align}
where $|+\rangle^{x}_{i}$ is the eigenstate of $X^{\,}_{i}$ with the
eigenvalue $+1$ and
\begin{equation}
G(\beta)\:=
\exp\left(\frac{\beta}{2}\sum_{j} Z^{\,}_{j}\,Z^{\,}_{j+1}\right)
\label{eq:1D scar operator}
\end{equation}
\end{subequations}
is annihilated by the operators $Q^{\,}_{i}(\beta)$ for all $i$. Therefore
$|\textrm{scar}(\beta)\rangle$ is an eigenstate of $H(\beta)$ with
eigenvalue 0.

That this eigenstate obeys 
area law entanglement entropy can be seen as follows.
The operators $Q^{\,}_{i}(\beta)$ are positive-semidefinite
definite, owing to the identity
$Q^{2}_{i}(\beta)=
2
\cosh
\left(\vphantom{\Bigg(}
\beta\,
\left(
Z^{\,}_{i-1}\,Z^{\,}_{i}
+
Z^{\,}_{i}\,Z^{\,}_{i+1}
\right)
\right)\,
Q^{\,}_{i} (\beta)$. Therefore,
$|\textrm{scar} \rangle$ is the ground state of another (local)
Hamiltonian, $\widehat{H}(\beta)\:=\sum_{i} |\alpha^{\,}_{i}|
Q^{\,}_{i}(\beta)$. The spectrum of $\widehat{H}(0)$ has a gap between
its ground state and all excited states, a gap that remains for a
finite range of values of $\beta$. Therefore,
$|\textrm{scar}(\beta)\rangle$ obeys
area law entanglement entropy for a range of
$\beta$~\cite{hastings2007area}.
Alternatively, the area-law property of
$|\textrm{scar}(\beta)\rangle$
can be argued from the form of Eq.~\eqref{eq:1Dscar3a3b} for any $\beta$,
by noting that it can be represented by a quantum circuit of constant depth (independent of both $\beta$ and system size), applied
to a product state~\cite{eisert2010colloquium,hermanns2017entanglement}. 

\begin{figure}[t]
\centering
\includegraphics[width=0.46\textwidth]{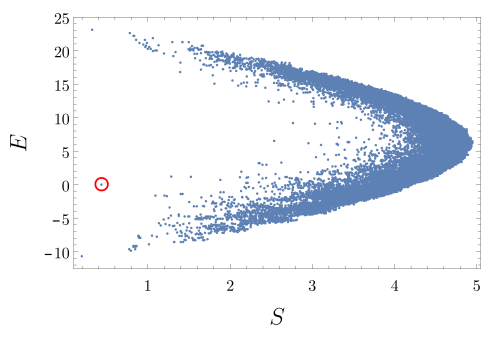}
\caption{
(Color online)
Entanglement entropy of the eigenstates of Hamiltonian
\eqref{eq:1D Hamiltonian} for a real-space bipartition of the
system into two equal halves. The parameters are set at $L=16$,
$\beta=0.5$, and $\alpha=0.3$.  The analytically obtained scar state
has $E=0$ (red circle) and is well-separated from the highly
entangled states.
        }
\label{fig:entanglement entropy}
\end{figure}

In Fig.~\ref{fig:entanglement entropy}, we present the entanglement
entropy for the different eigenstates of $H(\beta)$ for $\alpha=0.3$,
$\beta=0.5$ and $L=16$. Notice that the $E=0$ scar state is embedded
within highly entangled states.

\textit{1D: SPT cluster model ---} Consider a quantum spin-1/2 ring
with $2L$ sites. Odd and even sites are denoted by
$\mathrm{SL}^{\,}_{1}\:=\{1,3,\cdots,2L-1\}$ and
$\mathrm{SL}^{\,}_{2}\:=\{2,4,\cdots,2L\}$, respectively. For any
$\beta^{\,}_{\mathtt{a}}\geq0$ with $\mathtt{a}=1,2$, we define the
Hamiltonians
\begin{subequations}
\label{eq: def 1D Hmp}
\begin{align}
&
H^{\mathrm{1D}}\:=
H^{\mathrm{1D}}_{1}
+
H^{\mathrm{1D}}_{2},
\qquad
H^{\mathrm{1D}}_{\mathtt{a}}\:=
\sum_{j\in\mathrm{SL}^{\,}_{\mathtt{a}}}
\alpha^{\mathrm{1D}}_{\mathtt{a},j}\,
Q^{\mathrm{1D}}_{\mathtt{a},j},
\label{eq: def 1D Hmp a}
\\
&
\alpha_{\mathtt{a},j}^{\mathrm{1D}}\:=\alpha+(-1)^{\frac{j-\mathtt{a}}{2}},
\qquad
Q^{\mathrm{1D}}_{\mathtt{a},j}\:=
e^{
-\beta^{\,}_{\mathtt{a}}
\left(
X^{\,}_{j-1}
+
X^{\,}_{j+1}
\right)
  }
-
Z^{\,}_{j-1}\,
X^{\,}_{j}\,
Z^{\,}_{j+1}.
\label{eq: def 1D Hmp b}
\end{align}
\end{subequations} 
Note that $[H^{\mathrm{1D}}_{1},H^{\mathrm{1D}}_{2}]=0$ for any
$\beta^{\,}_{1}$ and $\beta^{\,}_{2}$.  For $\beta^{\,}_{1}=\beta^{\,}_{2}=0$,
$H^{\mathrm{1D}}$ is exactly solvable and its ground state is a gapped
SPT state~\cite{gu2009tensor,pollmann2012symmetry}. Its
topological attributes originate from symmetry protected zero modes
that are localized at the two ends of an open chain when open
boundary conditions are imposed instead of periodic ones.  The
symmetry protecting the boundary states is
$\mathbb{Z}^{\,}_{2}\times\mathbb{Z}^{\,}_{2}$ as shown in the
Supplemental Material~\cite{Chen2013}.  Being gapped at
$\beta^{\,}_{1}=\beta^{\,}_{2}=0$, the SPT phase extends to
non-vanishing but sufficiently small $\beta^{\,}_{1}>0$ and
$\beta^{\,}_{2}>0$.
(See Ref.~\cite{Santos2015} for another deformation of 1D SPT Hamiltonians.)

The null state for $\beta^{\,}_{1}=\beta^{\,}_{2}=0$ is an eigenstate
of the $Z^{\,}_{i-1}\,X^{\,}_{i}\,Z^{\,}_{i+1}$ operators,
$i=1,\cdots,2L$, with eigevalue $+1$. We denote this state by
$|+,\cdots,+\rangle$. For $\beta^{\,}_{1}>0$ and $\beta^{\,}_{2}>0$,
the null state of Eq.~(\ref{eq: def 1D Hmp a}) is
\begin{subequations}
\label{eq: def scar state}
\begin{equation}
|\mathrm{scar}^{\mathrm{1D}}\rangle\:=
G^{\mathrm{1D}}_{1}\,
G^{\mathrm{1D}}_{2}\,
|+,\cdots,+\rangle,
\label{eq: def scar state a}
\end{equation}
obtained via a similarity transformation with
\begin{equation}
G^{\mathrm{1D}}_{\mathtt{a}}\:=
\exp
\left(
\frac{\beta^{\,}_{\mathtt{a}}}{2}
\sum\limits_{j\in\mathrm{SL}^{\,}_{\mathtt{a}}}
X^{\,}_{j-1}
\right).
\label{eq: def scar state b}
\end{equation}
\end{subequations}

It remains to be shown that the Hamiltonian is non-integrable. Since
the Hamiltonian is made up of two commuting pieces $H^{\mathrm{1D}}_{1}$
and $H^{\mathrm{1D}}_{2}$, one must show that each component alone is
non-integrable. We shall reduce the calculation of the energy level
statistics to the problem already solved for the topologically trivial
warm up example of the Hamiltonian $H(\beta)$ in Eq.~\eqref{eq:1D Hamiltonian}, presented
previously. The mapping is via a nonlocal unitary transformation
\begin{equation}
W\:=
\exp
\left(
\,\mathrm{i}\frac{\pi}{4}
\sum\limits_{j\in\mathrm{SL}^{\,}_{1}}
Z^{\,}_{j}\,Z^{\,}_{j+1}
-\mathrm{i}\frac{\pi}{4}
\sum\limits_{j\in\mathrm{SL}^{\,}_{2}}
Z^{\,}_{j}\,Z^{\,}_{j+1}
\right),
\label{eq:gauge-W}
\end{equation}
which maps
$Q^{\mathrm{1D}}_{\mathtt{a},j}$
into
$
\widetilde{Q}^{\mathrm{1D}}_{\mathtt{a},j}\:=
W\;Q^{\mathrm{1D}}_{\mathtt{a},j}\;W^{\dag}$
where
\begin{equation}
{\widetilde Q}^{\mathrm{1D}}_{\mathtt{a},j}=
e^{
-
\beta^{\,}_{\mathtt{a}}\,
\left(
Z^{\,}_{j-2}\, X^{\,}_{j-1}\, Z^{\,}_{j}
+
Z^{\,}_{j}\, X^{\,}_{j+1}\, Z^{\,}_{j+2}
\right)
  } 
-
X^{\,}_{j}.
\label{eq:Q-tilde-1D}
\end{equation}
The spectrum of $H^{\mathrm{1D}}_{\mathtt{a}}$ can be related to that
of $H$ by noticing that the operators $X^{\,}_i$ with $i\in
\mathrm{SL}^{\,}_{2}$ that appear in the exponentials in
Eq.~(\ref{eq:Q-tilde-1D}) have no dynamics within
$H^{\mathrm{1D}}_{1}$, and {\it vice versa}, the $X^{\,}_i$ with $i\in
\mathrm{SL}^{\,}_{1}$ have no dynamics within
$H^{\mathrm{1D}}_{2}$. For the purpose of obtaining the eigenvalues of
$H^{\mathrm{1D}}_{1}$, one can freeze the $X^{\,}_i, i\in
\mathrm{SL}^{\,}_{2}$; there are only two gauge inequivalent choices
depending on the $\mathbb{Z}^{\,}_{2}$ sector selected, i.e., the choice of
$\prod_{i\in\mathrm{SL}^{\,}_{2}} X^{\,}_i=\pm 1$. (This symmetry is one of
the two $\mathbb{Z}^{\,}_{2}$'s in the
$\mathbb{Z}^{\,}_{2}\times \mathbb{Z}^{\,}_{2}$.)
The spectrum of $H^{\mathrm{1D}}_{1}$ in the $+$ sector (equivalent to
fixing $X^{\,}_i=+1, i\in \mathrm{SL}^{\,}_{2}$) reduces to that of
$H$ that we studied previously. We thus conclude that the 1D SPT scar
from Eq.~\eqref{eq: def scar state a} is an exceptional state in the spectrum of a
non-integrable Hamiltonian
$H^{\mathrm{1D}}_{1}$+$H^{\mathrm{1D}}_{2}$.

\begin{figure}[t]
\begin{center}
\includegraphics[width=0.48\textwidth]{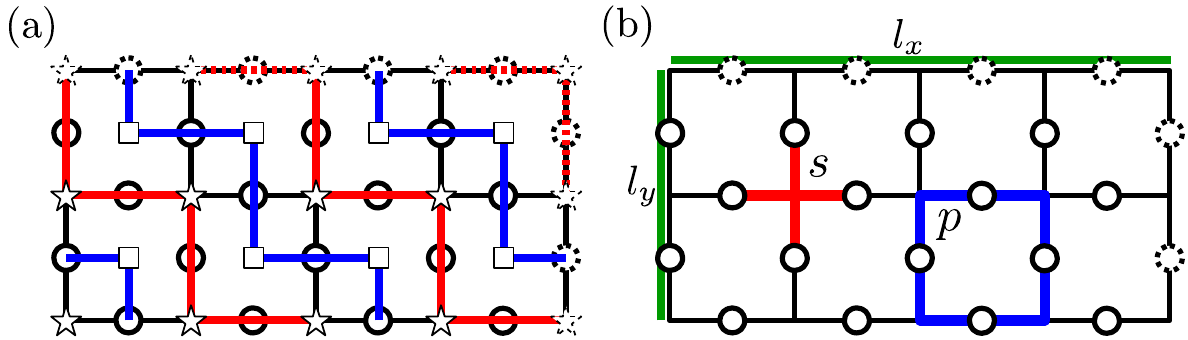}
\end{center}
\caption{
(Color online)
Example of a lattice structure of the 2D model. 
Dashed sites and lines 
are used to represent periodic boundary conditions. 
(a)
Starting from a ($N^{\,}_{x}\times N^{\,}_{y}=2\times4$)
square lattice $\Lambda^{\,}_{\star}$, we define the median and dual
lattices $\Lambda^{\,}_{\medcirc}$ and $\Lambda^{\,}_{\Box}$ in such a way that
sites of
$\Lambda^{\,}_{\star}$,
$\Lambda^{\,}_{\medcirc}$,
and $\Lambda^{\,}_{\Box}$ are represented by the symbols
$\bigstar$,
$\medcirc$,
and
$\square$,
respectively.
The red (blue) path $\mathcal{P}^{\,}_{1}$ ($\mathcal{P}^{\,}_{2}$)
along the bonds of $\Lambda^{\,}_{\star}$ ($\Lambda^{\,}_{\Box}$) goes
through all sites $\bigstar\in\Lambda^{\,}_{\star}$
($\square\in\Lambda^{\,}_{\Box}$) without intersecting itself.  (b)
The toric code assigns a local spin-1/2 degree of freedom to each site
$\medcirc$ of the median lattice $\Lambda^{\,}_{\medcirc}$. To each
site $\bigstar$ ($\square$) of the lattice $\Lambda^{\,}_{\star}$
($\Lambda^{\,}_{\Box}$), we assign the subset $s$ ($p$) consisting of
the 4 sites of $\Lambda^{\,}_{\medcirc}$ on the red cross (blue
square) at the site $\bigstar$ ($\square$) and define the star
(plaquette) operator $A^{\,}_{s} := \prod_{i\in s} X^{\,}_{i}$
($B^{\,}_{p}\, :=\prod_{i\in p} Z^{\,}_{i}$).  The two orthogonal
green lines are the ``electric'' paths $l^{\,}_{x}$ and $l^{\,}_{y}$
needed to define two Wilson loops
$W^{\,}_{\mu}:=\prod_{i\in l^{\,}_{\mu}\cap\Lambda^{\,}_{\medcirc}} Z^{\,}_{i}$
with $\mu=x,y$,
respectively.  }
\label{fig: Example lattice 2D model}
\end{figure}

\textit{Example in 2D: Toric code ---}
In 2D we study a lattice model derived from the toric code%
~\cite{KitaevToric2003}.
The Hamiltonian
$H^{\mathrm{2D}}\:=
H^{\mathrm{2D}}_{1}+H^{\mathrm{2D}}_{2}$
is defined by the pair of commuting operators
\begin{subequations}
\label{eq: def H1 and H2 in 2D}
\begin{align}
H^{\mathrm{2D}}_{1}&\:=
\sum_{s}\;
\alpha^{\,}_{s}
\left[\;
\exp
\left(
-\beta^{\,}_{1}
\sum_{i\in s \cap\mathcal{P}^{\,}_{1}}
Z^{\,}_{i}
\right)
-
A^{\,}_{s}
\;\right],
\label{eq: def H1 and H2 in 2Da}
\\
H^{\mathrm{2D}}_{2}&\:=
\sum_{p}\;
\alpha^{\,}_{p}
\left[\;
\exp
\left(
-\beta^{\,}_{2}
\sum_{i\in p \cap\mathcal{P}^{\,}_{2}}
X^{\,}_{i}
\right)
-
B^{\,}_{p}
\;\right],
\label{eq: def H1 and H2 in 2Db}
\end{align}
\end{subequations}
where $s$ labels a star and $p$ a plaquette (see
Fig.~\ref{fig: Example lattice 2D model}),
$A^{\,}_{s}=\prod_{i\in s} X^{\,}_{i}$ and
$B^{\,}_{p}=\prod_{i\in p} Z^{\,}_{i}$.
(Notice that $\beta^{\,}_{1,2}=0$
yields the usual toric code up
to an additive constant.) 
We define
$\alpha^{\,}_{s}\:=\alpha+(-1)^{\rho^{\,}_{s}}$
[$\alpha^{\,}_{p}\:=\alpha+(-1)^{\rho^{\,}_{p}}$]
such that $\rho^{\,}_{s}$ ($\rho^{\,}_{p}$)
is equal to 0 on one sublattice and 1 on the other sublattice of the lattice $\Lambda^{\,}_{\star}$ ($\Lambda^{\,}_{\square}$). 
Here, $\Lambda^{\,}_{\star}$ is the lattice formed by the centers of
all the stars, and $\Lambda^{\,}_{\Box}$ is the lattice formed by the centers 
of all the plaquettes. Our deformation of the toric code for
$\beta^{\,}_{1,2}\ne0$ uses the paths $\mathcal{P}^{\,}_{\mathtt{1}}$ and
$\mathcal{P}^{\,}_{\mathtt{2}}$, on $\Lambda^{\,}_{\star}$ and $\Lambda^{\,}_{\Box}$,
respectively. These paths are connected, non-intersecting, and chosen such that
all the spins are on either of the two paths. (An example of such paths
$\mathcal{P}^{\,}_{\mathtt{1,2}}$ is presented in
Fig.\ \ref{fig: Example lattice 2D model},
and in the Supplemental Material we give
further examples.) 
These conditions on
$\mathcal{P}^{\,}_{\mathtt{1,2}}$ guarantee that
(a) $[H^{\mathrm{2D}}_{1},H^{\mathrm{2D}}_{2}]=0$,
(b) there is no further integral of motion besides $H^{\mathrm{2D}}_{1}$ or
$H^{\mathrm{2D}}_{2}$ as well as space group symmetries, 
and (c) the spectrum of $H^{\textrm{2D}}_{1}$ alone is equal to that of
$H^{\textrm{1D}}_{1}$ for a path $\mathcal{P}^{\,}_{1}$ of length $L$ (up to
exact degeneracies due to a different number of integrals of motion
in 1D and 2D).
To obtain (c), one notes that $Z^{\,}_i$ for spins not in
$\mathcal{P}^{\,}_{\mathtt{2}}$ are integrals of motion of
$H^{\mathrm{2D}}_{2}$. Replacing them by their eigenvalue $\pm1$
reduces $H^{\mathrm{2D}}_{2}$ to the form of $H^{\mathrm{1D}}_{2}$ for
an appropriate choice of its integrals of motion $X^{\,}_j$ for $j\in
\mathrm{SL}_{2}$ in Eq.~\eqref{eq: def 1D Hmp b}, upon labeling the
spins along $\mathcal{P}^{\,}_{\mathtt{2}}$ in the order of the 1D
chain. We conclude that the level statistics of $H^{\textrm{2D}}_{1}$
and $H^{\textrm{1D}}_{1}$ are identical up to exact
degeneracies. Hence the numerical evidence for the non-integrability
of $H^{\textrm{1D}}_{1}$ directly carries over to
$H^{\textrm{2D}}_{1}$.  In our model, the extensive symmetries at
$\beta^{\,}_{1}=\beta^{\,}_{2}=0$ arising from
$[A^{\,}_{s},B^{\,}_{p}]=0$ are lifted when $\beta^{\,}_{1,2}\neq0$
(in which case $H^{\mathrm{2D}}_{1,2}$ are no longer sums of commuting
projectors).

The scar states are built as follows. Because $A^{\,}_{s}$ and
$B^{\,}_{p}$ square to unity and satisfy
$\prod_{s}A^{\,}_{s}=\prod_{p}B^{\,}_{p}=\openone$, we can build a
vector $\bm{\lambda}\in\{-,+\}^{2N^{\,}_{x}\,N^{\,}_{y}-2}$ out of the
distinct eigenvalues of $(N^{\,}_{x}\,N^{\,}_{y}-1)$ independent
$A^{\,}_{s}$'s and $(N^{\,}_{x}\,N^{\,}_{y}-1)$ independent
$B^{\,}_{p}$'s to label an orthogonal basis $|\bm{\lambda}\rangle$ of
a $2^{2N^{\,}_{x}\,N^{\,}_{y}-2}$-dimensional subspace of the
$2^{2N^{\,}_{x}\,N^{\,}_{y}}$-dimensional Hilbert space on which
$H^{\mathrm{2D}}$ acts. To complete the basis of the Hilbert
space, we use the eigenstates $|\bm{\omega}\rangle$ with the
eigenvalues $\bm{\omega}\equiv(\omega^{\,}_{x}=\pm,\omega^{\,}_{y}=\pm)$
of the pair of Wilson-loop operators $W^{\,}_{\mu}$ with $\mu=x,y$
defined in Fig.~\ref{fig: Example lattice 2D model}. 
The following four scar states (one in each of the
4 topological sectors) are eigenstates of $H^{\mathrm{2D}}$ with
the eigenvalues $E=0$:
\begin{subequations}
\begin{align}
&
|\mathrm{scar}^{\textrm{2D}};\bm{\omega}\rangle\:=
G^{\mathrm{2D}}_{1}\,
G^{\mathrm{2D}}_{2}\,
|+,\cdots,+;\bm{\omega}\rangle,
\\
&G^{\mathrm{2D}}_{1}\:=
\exp\left(
\frac{\beta^{\,}_{1}}{2}
\sum_{i\in\mathcal{P}^{\,}_{1}}
Z^{\,}_{i}
\right)
\;,
\quad
G^{\mathrm{2D}}_{1}\:=
\exp\left(
\frac{\beta^{\,}_{2}}{2}
\sum_{i\in\mathcal{P}^{\,}_{2}}
X^{\,}_{i}
\right).
\end{align}
\end{subequations}

\begin{figure}[t]
\begin{center}
\includegraphics[width=0.30\textwidth]{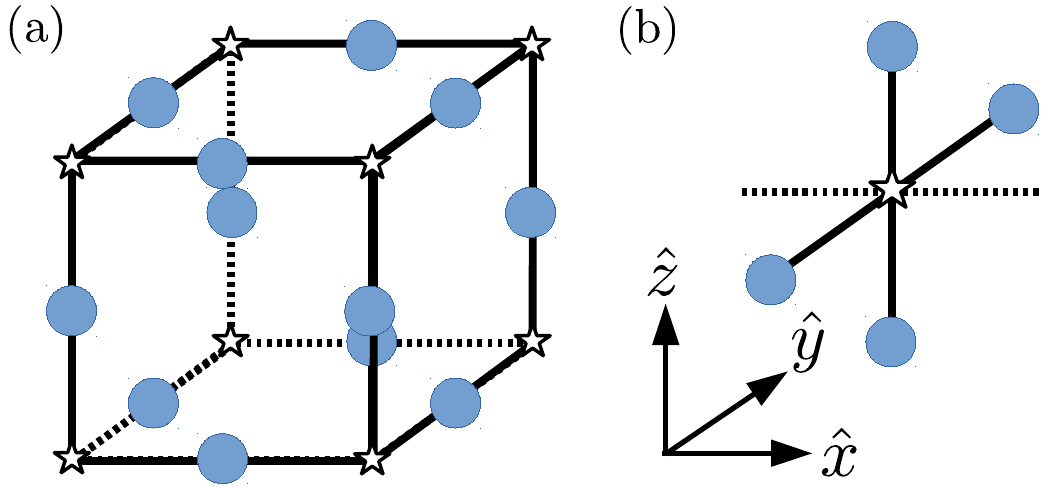}
\end{center}
\caption{\label{fig: X-cube}
(Color online) The notation
$\Lambda^{\,}_{\star}$,
$\Lambda^{\,}_{\medcirc}$,
and $\Lambda^{\,}_{\Box}$
of Fig.\ \ref{fig: Example lattice 2D model} 
becomes
$\Lambda^{\,}_{\star}$,
$\Lambda^{\,}_{\medcirc}$,
and $\Lambda^{\,}_{\textrm{\cube}}$,
where $\Lambda^{\,}_{\star}$ denotes the cubic lattice,
$\Lambda^{\,}_{\medcirc}$ its median lattice, and
$\Lambda^{\,}_{\textrm{\cube}}$ its dual lattice.
(a)
The elementary unit cell $\textrm{\cube}$ of $\Lambda^{\,}_{\star}$
is cubic. Spin-1/2 degrees of freedom represented by $\medcirc$
are located on its mid-bonds.
The 12 $\medcirc$s on the bonds of a $\textrm{\cube}$ define a subset
$c\subset\Lambda^{\,}_{\medcirc}$.
The corners of $\textrm{\cube}$ define sites $\bigstar$ of $\Lambda^{\,}_{\star}$.
The center of $\textrm{\cube}$ defines a site from $\Lambda^{\,}_{\textrm{\cube}}$.
For any such $\textrm{\cube}$, we define
$B^{\,}_{c}$ by taking the product of all 12 Pauli matrices $Z^{\,}_{i}$
from the neighboring bonds with $i\in c\cap\Lambda^{\,}_{\medcirc}$.
(b)
The center of a cross $\boldplus$ joining its 4 nearest-neighbor sites from
$\Lambda^{\,}_{\medcirc}$ defines a site from $\Lambda^{\,}_{\star}$ and
the subset $s\subset\Lambda^{\,}_{\medcirc}$.
There are three oriented crosses for any site from $\Lambda^{\,}_{\star}$.
They are in one-to-one correspondence with the three oriented planes
in the Cartesian coordinates of $\mathbb{R}^{3}$.
For any such oriented cross, we define
$A^{\,}_{s}$ by taking the product of all four Pauli matrices
$X^{\,}_{i}$ with $i\in s$.
        }
\end{figure}

\textit{3D Example: $X$-cube model} ---
Our 2D construction can 
be extended in a straightforward way to 3D
toric code-type Hamiltonians~\cite{hamma2005}. Here, we derive scar
states for the slightly more exotic fracton topological order, which
only arises in three or more dimensions%
~\cite{Chamon2005,Castelnovo2010,SagarFracton2015,Haah2011,SagarXCube2016}.
Fracton phases carry excitations which are (at least partially)
immobile in that they cannot be moved infinitesimally by applying
local operators.  In addition, they can support topological ground
state degeneracies that scale exponentially in the system size. Here,
we introduce a Hamiltonian based on the X-cube
model~\cite{SagarXCube2016}, which supports fracton topological order
in its ground state, to construct a set of 3D scar states with the
same exponential degeneracy. The Hamiltonian $H^{\mathrm{3D}}\:=
H^{\mathrm{3D}}_{1}+H^{\mathrm{3D}}_{2}$ is, once again, defined by
the pair of commuting operators
\begin{subequations}
\begin{align}
H^{\mathrm{3D}}_{1}&\:=
\sum_{s}\;
\alpha^{\,}_{s}
\left[\;
\exp
\left(
-\beta^{\,}_{1}
\sum_{i\in s \cap\mathcal{P}^{\,}_{1}}
Z^{\,}_{i}
\right)
-
A^{\,}_{s}
\;\right],
\label{eq:15fdsaf13D}
\\
H^{\mathrm{3D}}_{2}&\:=
\sum_{c}\;
\alpha^{\,}_{c}
\left[\;
\exp
\left(
-\beta^{\,}_{2}
\sum_{i\in c \cap\mathcal{P}^{\,}_{2}}
X^{\,}_{i}
\right)
-
B^{\,}_{c}
\;\right],
\label{eq:15fdsaf23D}
\end{align}
\end{subequations}
where $s$ labels a star and $c$ a cube (see Fig.\ \ref{fig: X-cube}),
$A^{\,}_{s}=\prod_{i\in s} X^{\,}_{i}$ and $B^{\,}_{c}=\prod_{i\in c}
Z^{\,}_{i}$. (Notice that $\beta^{\,}_{1,2}=0$ yields the usual X-cube
model up to a constant.) 
We define $\alpha^{\,}_{s}\:=\alpha+(-1)^{\rho^{\,}_{s}}$
($\alpha^{\,}_{c}\:=\alpha+(-1)^{\rho^{\,}_{c}}$)
analogously to that in the 2D model,
such that $\rho^{\,}_{s}$ ($\rho^{\,}_{c}$) is equal to 0 on one
sublattice and 1 on the other sublattice of the lattice
$\Lambda^{\,}_{\star}$ ($\Lambda^{\,}_{\textrm{\cube}}$).
The paths $\mathcal{P}^{\,}_{\mathtt{1}}$ and
$\mathcal{P}^{\,}_{\mathtt{2}}$ are defined on $\Lambda^{\,}_{\star}$
and $\Lambda^{\,}_{\textrm{\cube}}$, respectively, and they obey the
same conditions as in the 2D construction. These conditions guarantee
that $[H^{\mathrm{3D}}_{1},H^{\mathrm{3D}}_{2}]=0$ for any
$\beta^{\,}_{1,2}$, while lifting the extensive symmetries at
$\beta^{\,}_{1}=\beta^{\,}_{2}=0$ arising from
$\left[A^{\,}_{s},B^{\,}_{c}\right]=0$ because $H^{\textrm{3D}}_{1,2}$
are no longer sums of commuting projectors.

The Hilbert space for a cubic lattice of linear size $L$ is
$2^{3L^{3}}$-dimensional (there are $L^{3}$ sites in $\Lambda^{\,}_{\star}$ and
$3L^{3}$ in $\Lambda^{\,}_{\medcirc}$). The counting of independent
stars and cubes delivers the vector
$\bm{\lambda}\in\{-,+\}^{3L^{3}-6L+3}$ of eigenvalues. These
quantum numbers are complemented by the sub-extensive vector
$\bm{\zeta}\in\{-,+\}^{6L-3}$ of topological quantum numbers.  The
number of scar states that are eigenstates of $H^{\mathrm{3D}}$
with the eigenenergy $E=0$ thus grows sub-extensively with the linear
size $L$ of $\Lambda^{\,}_{\star}$, and are written as
\begin{subequations}
\begin{align}
&
|\mathrm{scar}^{\textrm{3D}};\bm{\zeta}\rangle\:=
G^{\mathrm{3D}}_{1}\,
G^{\mathrm{3D}}_{2}\,
|+,\cdots,+;\bm{\zeta}\rangle,
\\
&G^{\mathrm{3D}}_{1}\:=
\exp\left(
\frac{\beta^{\,}_{1}}{2}
\sum_{i\in\mathcal{P}^{\,}_{1}}
Z^{\,}_{i}
\right)
\;,
\quad
G^{\mathrm{3D}}_{2}\:=
\exp\left(
\frac{\beta^{\,}_{2}}{2}
\sum_{i\in\mathcal{P}^{\,}_{2}}
X^{\,}_{i}
\right).
\end{align}
\end{subequations}

\textit{Conclusions} --- We proposed a scheme to analytically
construct highly excited states of non-integrable local Hamiltonians
with sub-volume-law entanglement entropy scaling that are embedded in
a dense spectrum of volume-law scaling states. We gave further examples of
constructions of scar states using stochastic matrix form
Hamiltonians~\cite{Henley1997,Castelnovo2005,ChamonCastelnovo2008}
with a notion of SPT or topological orders. This allowed us to construct
sets of degenerate scar states. Whether these degeneracies are
topological in that they carry a sense of protection against small
generic local perturbations is left as a problem for future work.

\section*{Acknowledgments}
The authors thank Nicolas Regnault for discussions and insightful comments on the manuscript.
TN and CCa thank Zlatko Papi\'c for fruitful
discussions. SO and TN were supported by the the Swiss National
Science Foundation (grant number: 200021\_{1}69061).  KC and TN were
supported by the European Unions Horizon 2020 research and innovation
program (ERC-StG-Neupert-757867-PARATOP). CCh was supported by the
U.S. Department of Energy (DOE), Division of Condensed Matter Physics
and Materials Science, under Contract No.~DE-FG02-06ER46316. 
CCa was supported in part by Engineering and Physical Sciences Research 
Council (EPSRC) Grants No.~EP/P034616/1 and No.~EP/M007065/1. 
CCh thanks the hospitality of the Pauli Center for Theoretical Studies at
ETH Z\"urich and the University of Z\"urich, where this work
was started.

\bibliography{Draft}
\clearpage

\begin{widetext}

\section{Supplemental Material}


\subsection{Construction of Hamiltonians containing null states}

Here we demonstrate the construction of Hamiltonians hosting null eigenstates starting from a
solvable model.

Consider first operators $A^{\,}_{s}$ satisfying
\begin{subequations}
\begin{align}
A^{2}_{s}=\openone,
\qquad
\left[A^{\,}_{s}\,,\,A^{\,}_{s'}\right]=0, \quad \forall s,s',
\end{align}
where the $s$ label bounded regions in space, for instance any finite subset
of sites from a lattice. The notion of locality is tied to the fact that
the region on which the operators act nontrivially 
is bounded. More precisely, for two sites $i,j\in s$, the
distance between the sites is bounded, $|i-j|<d^{\,}_{s}$, where $d^{\,}_{s}$ is
the finite ``diameter'' of the region $s$.
Notice that the operators $\openone - A^{\,}_{s}$ are commuting projectors.
Second, we define 
\begin{align}
M\:=
\sum_{i}
O^{\,}_{i},
\qquad
M^{\,}_{s}\:=
\sum_{i\in s}
O^{\,}_{i},
\qquad
\overline{M}^{\,}_{s}\:=
\sum_{i\notin s}
O^{\,}_{i},
\end{align}
where the operators $O^{\,}_{i}$ need not just act at one site $i$, but on
a bounded subset of sites centered around $i$. The operators $O^{\,}_{i}$
are chosen to be Hermitian and to commute,
\begin{align}
  \left[O_{i}\,,\,O_{j}\right]=0,
\qquad \forall i,j
\, , 
\end{align}
as well as such that
\begin{align}
\left\{A^{\,}_{s}\,,\,M_{s}\right\}=0,
\qquad
\left[A^{\,}_{s}\,,\,\overline{M}^{\,}_{s}\right]=0,
\qquad
\forall s.
\end{align}
\end{subequations}
(Notice that if $O^{\,}_{i}$ contains exclusively operators at site $i$,
that $\left[A^{\,}_{s}\,,\,\overline{M}_{s}\right]=0$
follows trivially from the fact
that no common site belongs to $s$ and its complement.)
Third, we define
\begin{subequations}
\begin{align}
F^{\,}_{s}\:=&\;
e^{+\frac{1}{2}\beta\,M}\;\left(\openone-A^{\,}_{s}\right)\;e^{-\frac{1}{2}\beta\,M}
\nonumber\\
= &\;
\openone-e^{+\beta\,M^{\,}_{s}}\;A^{\,}_{s}
\nonumber\\
= &\;
e^{+\beta\,M^{\,}_{s}}\;\left(e^{-\beta\,M^{\,}_{s}}\;-A^{\,}_{s}\right),
\end{align}
and
\begin{align}
Q^{\,}_{s}\:=
e^{-\beta\,M_{s}}\; - A^{\,}_{s}.
\end{align}
\end{subequations}
Notice that $Q^{\,}_{s}$ is Hermitian, while $F^{\,}_{s}$ is not. They are
related by
\begin{align}
Q^{\,}_{s} = e^{-\beta\,M^{\,}_{s}}\;F^{\,}_{s}\;.
\end{align}
In addition to being Hermitian, $Q^{\,}_{s}$ is local, because $A^{\,}_{s}$ is
local and the exponential of the local operator $M^{\,}_{s}$ is also
local; and it is positive-semidefinite, as can be
inferred by squaring it,
\begin{align}
Q^{2}_{s}=
2\,\cosh(\beta\,M^{\,}_{s})\;Q^{\,}_{s}\;,
\end{align}
and observing that $\cosh(\beta\,M^{\,}_{s})$ is positive-definite.

We shall now construct a common null state to all the $Q^{\,}_{s}$
operators.

First, notice that the state
\begin{align}
|\Psi^{\,}_{0}\rangle\:=
\prod_{s'}
(\openone+A^{\,}_{s'})\;|\Omega\rangle
\end{align}
is annihilated by $(\openone-A^{\,}_{s})$, for all $s$, for
\begin{align}
(\openone - A^{\,}_{s})\;|\Psi^{\,}_{0}\rangle=&\,
(\openone - A^{\,}_{s})\;\prod_{s'} (\openone + A^{\,}_{s'})\;|\Omega\rangle
\nonumber\\
=&\,
(\openone - A^{\,}_{s})\,
(\openone + A^{\,}_{s})\;
\prod_{s'\ne s}
(\openone + A^{\,}_{s'})\;|\Omega\rangle
\nonumber\\
=&\,
(\openone - A^{2}_{s})\;
\prod_{s'\ne s} (\openone + A^{\,}_{s'})\;|\Omega\rangle
\nonumber\\
=&\,
0\;,
\end{align}
where we used the fact that $A^{2}_{s}=\openone$. The state $|\Omega\rangle$ 
is arbitrary, as long as it is not annihilated by the projectors
$(\openone + A^{\,}_{s'})$.

Second, let
\begin{align}
|\Psi^{\,}_{\beta}\rangle\:=
e^{+\frac{1}{2}\beta\,M}\;|\Psi^{\,}_{0}\rangle\;.
\end{align}
It follows that, for any $s$,
\begin{align}
F^{\,}_{s}\;|\Psi^{\,}_{\beta}\rangle=&\,
e^{+\frac{1}{2}\beta\,M}\;(\openone - A^{\,}_{s})\;e^{-\frac{1}{2}\beta\,M}\;
e^{+\frac{1}{2} \beta\,M}\;|\Psi^{\,}_{0}\rangle
\nonumber\\
=&\,
e^{+\frac{1}{2}\beta\,M}\;(\openone - A^{\,}_{s})\;|\Psi^{\,}_{0}\rangle
\nonumber\\
=&\,
0\;,
\end{align}
and consequently
\begin{align}
Q^{\,}_{s}\;|\Psi^{\,}_{\beta}\rangle=
e^{-\beta\,M^{\,}_{s}}\;F^{\,}_{s}\;|\Psi^{\,}_{\beta}\rangle=
0\;.
\end{align}

Therefore, the state $|\Psi^{\,}_{\beta}\rangle$ is a common null state of
all the local operators $Q^{\,}_{s}$, and also of any local Hamiltonian
written as a weighted sum of the $Q^{\,}_{s}$, say
\begin{equation}
H(\beta)\:=\sum_{s}\alpha^{\,}_{s}\,Q^{\,}_{s},
\end{equation}
for any weights $\alpha^{\,}_{s}\in\mathbb{R}$. 
In Eq.~\eqref{eq:1D Hamiltonian},
we chose, in place of $A^{\,}_{s}$ and $M^{\,}_{s}$,
$X^{\,}_{i}$ and $-\beta\,(Z^{\,}_{i-1}\,Z^{\,}_{i}+Z^{\,}_{i}\,Z^{\,}_{i+1})$,
respectively.

\subsection{Symmetries in 1D}

One finds the commutation relations
\begin{equation}
\left[
H^{\mathrm{1D}}_{1},
H^{\mathrm{1D}}_{2}
\right]=
\left[
H^{\mathrm{1D}},
H^{\mathrm{1D}}_{\mathtt{a}}
\right]=0,
\qquad
\mathtt{a}=1,2.
\end{equation}
Therefore, $H^{\mathrm{1D}}_{1}$,
$H^{\mathrm{1D}}_{2}$,
and
$H^{\mathrm{1D}}$ 
can be diagonalized simultaneously.

\textit{Translation symmetry:}
$H^{\mathrm{1D}}_{1}$,
$H^{\mathrm{1D}}_{2}$,
and
$H^{\mathrm{1D}}$
are each invariant under the translations
\begin{equation}
i\mapsto i+2n,\qquad i=1,\cdots,2L,\qquad n\in\mathbb{Z}. 
\label{eq: def translation symmetry supp}
\end{equation}
Hence, 
$H^{\mathrm{1D}}_{1}$,
$H^{\mathrm{1D}}_{2}$,
and
$H^{\mathrm{1D}}$
can be simultaneously diagonalized with the Hermitian generator of the
unitary operators representing the transformations
(\ref{eq: def translation symmetry supp}),
i.e., the momentum operator associated to the sublattice $\mathrm{SL}^{\,}_{1}$,
say. 

\textit{Inversion symmetry:}
For any site $j\in\mathrm{SL}^{\,}_{1}$,
$H^{\mathrm{1D}}_{1}$
is invariant under the inversion
\begin{equation}
i\mapsto i-2(i-j),\qquad i=1,\cdots,2L.
\label{eq: inversion H1 supp}
\end{equation}
For any site $j\in\mathrm{SL}^{\,}_{2}$,
$H^{\mathrm{1D}}_{2}$
is invariant under the inversion
\begin{equation}
i\mapsto i-2(i-j),\qquad i=1,\cdots,2L.
\label{eq: inversion H2 supp}
\end{equation}
Hence,
$H^{\mathrm{1D}}$
has the $\mathbb{Z}^{\,}_{2}\times\mathbb{Z}^{\,}_{2}$ symmetry that is
generated by the two independent involutive unitary transformations
(\ref{eq: inversion H1 supp})
and
(\ref{eq: inversion H2 supp}).
This is to say that
$H^{\mathrm{1D}}_{1}$,
$H^{\mathrm{1D}}_{2}$,
and
$H^{\mathrm{1D}}$
are invariant under any inversion of the ring that leaves one site of the
ring unchanged.

\textit{Two independent involutive symmetries:}
Hamiltonian
$H^{\mathrm{1D}}_{1}$
is invariant under the involutive unitary transformation
\begin{subequations}
\label{eq: Two independent involutive internal symmetries supp}
\begin{equation}
Z^{\,}_{j}\mapsto
U^{\,}_{2}\,
Z^{\,}_{j}\,
U^{\,}_{2}=
-Z^{\,}_{j},
\qquad
j\in\mathrm{SL}^{\,}_{2},
\qquad
U^{\,}_{2}\:=\prod_{k\in\mathrm{SL}^{\,}_{2}} X^{\,}_{k}=U^{\dag}_{2},
\label{eq: involutive unitary transformation H1 supp}
\end{equation}
that acts trivially on the sites of the ring.
Hamiltonian
$H^{\mathrm{1D}}_{2}$
is invariant under the involutive unitary transformation
\begin{equation}
Z^{\,}_{j}\mapsto
U^{\,}_{1}\,
Z^{\,}_{j}\,
U^{\,}_{1}=
-Z^{\,}_{j},
\qquad
j\in\mathrm{SL}^{\,}_{1},
\qquad
U^{\,}_{1}\:=\prod_{k\in\mathrm{SL}^{\,}_{1}} X^{\,}_{k}=U^{\dag}_{1},
\label{eq: involutive unitary transformation H2 supp}
\end{equation}
\end{subequations}
that acts trivially on the sites of the ring.
Hence,
$H^{\mathrm{1D}}$
has the $\mathbb{Z}^{\,}_{2}\times\mathbb{Z}^{\,}_{2}$ symmetry that is
generated by the two independent involutive unitary transformations
(\ref{eq: involutive unitary transformation H1 supp})
and
(\ref{eq: involutive unitary transformation H2 supp}).

\subsection{A local unitary transformation in 1D}

We verify the transformation law
\begin{equation}
Q^{\mathrm{1D}}_{\mathtt{a},j}\mapsto
{\widetilde{Q}}^{\mathrm{1D}}_{\mathtt{a},j} =
W\;Q^{\mathrm{1D}}_{\mathtt{a},j}\;W^{\dag},
\end{equation}
with
${\widetilde{Q}}^{\mathrm{1D}}_{\mathtt{a},j}$ and $W$ defined in
Eq.~\eqref{eq:Q-tilde-1D} and Eq.~\eqref{eq:gauge-W}, respectively.
To this end, it suffices to prove the identity
\begin{equation}
W\; X^{\,}_{i}\; W^{\dagger}=Z^{\,}_{i-1} X^{\,}_{i}\,Z^{\,}_{i+1}, \qquad\forall i.
\label{eq: WXW=ZXZ}
\end{equation}

The terms in the exponent of $W$ that do not contain $X^{\,}_{i}$ do
not contribute to the transformation, i.e.,
\begin{equation} 
W\; X^{\,}_{i}\; W^{\dag}= 
e^{
\pm \mathrm{i} \frac{\pi}{4}\, Z^{\,}_{i-1}\, Z^{\,}_{i}
\mp \mathrm{i} \frac{\pi}{4}\, Z^{\,}_{i}\, Z^{\,}_{i+1}
  }\;
X^{\,}_{i}\;
e^{
\mp \mathrm{i} \frac{\pi}{4}\, Z^{\,}_{i-1}\, Z^{\,}_{i}
\pm \mathrm{i} \frac{\pi}{4}\, Z^{\,}_{i}\, Z^{\,}_{i+1}
  }=
X^{\,}_{i}\,
e^{
\mp \mathrm{i} \frac{\pi}{2}\, Z^{\,}_{i-1}\, Z^{\,}_{i}
\pm \mathrm{i} \frac{\pi}{2} Z^{\,}_{i}\, Z^{\,}_{i+1}
  },
\end{equation}
where $\pm=+$, $\mp=-$ for $i\in\mathrm{SL}^{\,}_{1}$,
and vice versa for $i\in\mathrm{SL}^{\,}_{2}$.
Using additional relations
\begin{equation}
e^{\mp \textrm{i} \frac{\pi}{2} Z^{\,}_{i-1}\, Z^{\,}_{i} } = \mp \textrm{i}\, Z^{\,}_{i-1}\, Z^{\,}_{i}, \quad
e^{\pm \textrm{i} \frac{\pi}{2} Z^{\,}_{i}\, Z^{\,}_{i+1} } = \pm \textrm{i}\, Z^{\,}_{i}\, Z^{\,}_{i+1},
\end{equation}
one acquires the identity in Eq.~\eqref{eq: WXW=ZXZ}.

\subsection{Open boundary conditions in 1D}

Using the notation introduced
in Eq.~(\ref{eq: def 1D Hmp}),
we define the Hamiltonian
\begin{equation}
H^{\mathrm{1D}}_{\mathrm{OBC}}\:=
H^{\mathrm{1D}}_{1,\mathrm{OBC}}
+
H^{\mathrm{1D}}_{2,\mathrm{OBC}},
\qquad
H^{\mathrm{1D}}_{1,\mathrm{OBC}}\:= 
H^{\mathrm{1D}}_{1}
-
Q^{\mathrm{1D}}_{1,1},
\qquad
H^{\mathrm{1D}}_{2,\mathrm{OBC}}\:= 
H^{\mathrm{1D}}_{2}
-
Q^{\mathrm{1D}}_{2,2L}.
\end{equation}
By inspection of the explicit representations
\begin{equation}
H^{\mathrm{1D}}_{1,\mathrm{OBC}}=
\sum_{j=1}^{L-1}
e^{
-
\beta^{\,}_{1}
\left(
X^{\,}_{2j}
+
X^{\,}_{2j+2}
\right)
  }
-
Z^{\,}_{2j}\,
X^{\,}_{2j+1}\,
Z^{\,}_{2j+2},
\qquad
H^{\mathrm{1D}}_{2,\mathrm{OBC}}=
\sum_{j=1}^{L-1}
e^{
-
\beta^{\,}_{2}
\left(
X^{\,}_{2j-1}
+
X^{\,}_{2j+1}
\right)
  }
-
Z^{\,}_{2j-1}\,
X^{\,}_{2j}\,
Z^{\,}_{2j+1},
\end{equation}
we observe that $\Lambda^{\textrm{OBC}}_{1} \:=
X^{\,}_{1}\,Z^{\,}_{2}$ and $\Lambda^{\textrm{OBC}}_{2L} \:=
Z^{\,}_{2L-1}\,X^{\,}_{2L}$ obey the vanishing commutation relations
\begin{equation}
\begin{split}
&
\left[\Lambda^{\textrm{OBC}}_{1},H^{\mathrm{1D}}_{1,\mathrm{OBC}}\right]=
\left[\Lambda^{\textrm{OBC}}_{1},Z^{\,}_{i-1}\,X^{\,}_{i}\,Z^{\,}_{i+1}\right]=0,
\qquad i=3,\cdots,2L-1,
\\
&
\left[\Lambda^{\textrm{OBC}}_{2L},H^{\mathrm{1D}}_{2,\mathrm{OBC}}\right]=
\left[\Lambda^{\textrm{OBC}}_{2L},Z^{\,}_{i-1}\,X^{\,}_{i}\,Z^{\,}_{i+1}\right]=0,
\qquad i=2,\cdots,2L-2.
\end{split}
\end{equation}
The two vanishing anticommutators
\begin{equation}
\left\{\Lambda^{\mathrm{OBC}}_{1},U_{2}\right\}=
\left\{\Lambda^{\mathrm{OBC}}_{2L},U_{1}\right\}=0,
\end{equation}
along with the fact that $\Lambda^{\mathrm{OBC}}_{1}$,
$\Lambda^{\mathrm{OBC}}_{2}$ and the Hermitian operator
$U^{\,}_{\mathtt{a}}\equiv\prod_{j\in\mathrm{SL}^{\,}_{\mathtt{a}}}X^{\,}_{j}$
defined in Eq.~(\ref{eq: Two independent involutive internal symmetries supp})
commute with $H^{\mathrm{1D}}_{\mathrm{OBC}}$,
imply that every eigenspace of $H^{\mathrm{1D}}_{\mathrm{OBC}}$,
including the one of the scar state, is at least four-fold degenerate,
and the quadruplet of states can be labelled by the eigenvalues of
$\Lambda^{\mathrm{OBC}}_{1}$ and $\Lambda^{\mathrm{OBC}}_{2L}$.
The degeneracy is protected by the symmetries U1 and U2. Since 
$\Lambda^{\mathrm{OBC}}_{1}$ and $\Lambda^{\mathrm{OBC}}_{2L}$ are local 
operators at the end of the chain, the Hamiltonian is in an SPT phase.

\begin{figure}[t]
\begin{center}
\includegraphics[width=0.7\textwidth]{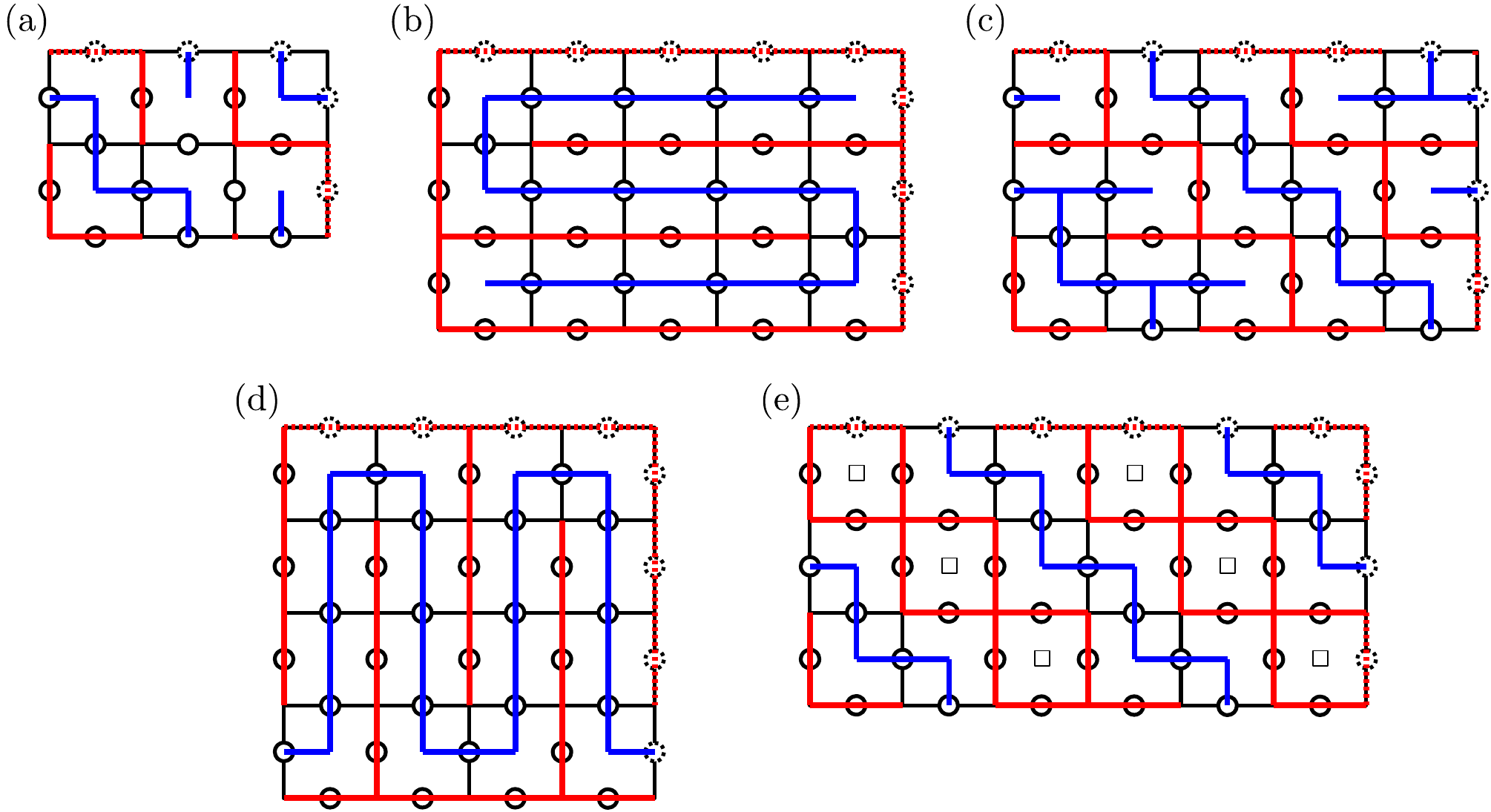}
\end{center}
\caption{
Exemples of lattice structures for the 2D model. 
Dashed sites and lines are used to represent periodic boundary conditions. 
Any path $\mathcal{P}^{\,}_{1}$ that is colored in red starts
and ends by definition
on the sites of the lattice $\Lambda^{\,}_{\star}$.
Any path $\mathcal{P}^{\,}_{2}$ colored in blue starts and ends by definition
on the sites of the dual lattice $\Lambda^{\,}_{\Box}$.
The spin degrees of freedom are located on the sites of the median
lattice $\Lambda^{\,}_{\medcirc}$ denoted by open circles.
(a)--(d)
Example of the path $\mathcal{P}^{\,}_{1}$ colored in red
and the path $\mathcal{P}^{\,}_{2}$ colored in blue for a square lattice
of given aspect ratio. 
Only the sites $i$ of $\Lambda^{\,}_{\medcirc}$
represented by open circles are shown.
With this choice for the paths
$\mathcal{P}^{\,}_{1}$and $\mathcal{P}^{\,}_{2}$,
the condition $\beta^{\,}_{1},\beta^{\,}_{2}>0$
is sufficient to guarantee that the sum over $s$ in 
$H^{\mathrm{2D}}_{1}$ (the sum over $p$ in $H^{\mathrm{2D}}_{2}$) 
can never be arranged into the sum of two non-vanishing
Hermitian operators that commute pairwise and commute with
$H^{\mathrm{2D}}_{2}$ ($H^{\mathrm{2D}}_{1}$). 
(e) The choice made for the path $\mathcal{P}^{\,}_{1}$ colored in red
and the path $\mathcal{P}^{\,}_{2}$ colored in blue fails to guarantee
that the sum in $H^{\mathrm{2D}}_{\mathtt{a}}$
can be arranged into the sum of two non-vanishing
Hermitian operators that commute pairwise and with
$H^{\mathrm{2D}}_{\bar{\mathtt{a}}}$
when $\beta^{\,}_{\mathtt{a}},\beta^{\,}_{\bar{\mathtt{a}}}>0$.
Indeed, of all Hermitian operators $B^{\,}_{p}$ entering $H^{\mathrm{2D}}_{2}$,
those sites from the dual lattice $\Lambda^{\,}_{\Box}$
that are identified by the symbol $\Box$
are not traversed by $\mathcal{P}^{\,}_{2}$.
They give a set of operators $\{B^{\,}_{\Box}\}$, whereby
$B^{\,}_{\Box}$ commutes with both $H^{\mathrm{2D}}_{1}$ and $H^{\mathrm{2D}}_{2}$.
       }
\label{fig: notrivial examples path}
\end{figure}

\subsection{Examples of paths $\mathcal{P}^{\,}_{1}$ and $\mathcal{P}^{\,}_{2}$
in 2D}

For convenience, we recall that we introduced the pair of Hamiltonians
\begin{align}
H^{\mathrm{2D}}_{1}\:=
\sum_{s}\;
\left[
\exp
\left(
-\beta^{\,}_{1}
\sum_{i\in s \cap\mathcal{P}^{\,}_{1}}
Z^{\,}_{i}
\right)
-
A^{\,}_{s}
\right],
\qquad
A^{\,}_{s}\:=\prod_{i\in s} X^{\,}_{i},
\qquad
H^{\mathrm{2D}}_{2}&\:=
\sum_{p}
\left[
\exp
\left(
-\beta^{\,}_{2}
\sum_{i\in p \cap\mathcal{P}^{\,}_{2}}
X^{\,}_{i}
\right)
-
B^{\,}_{p}
\right],
\qquad
B^{\,}_{p}\:=\prod_{i\in p} Z^{\,}_{i},
\end{align}
in Eq.\ (\ref{eq: def H1 and H2 in 2D}).
The definition of the paths
$\mathcal{P}^{\,}_{1}$ and $\mathcal{P}^{\,}_{2}$
was given below Eq.\ (\ref{eq: def H1 and H2 in 2D}).
An example for the choice of paths $\mathcal{P}^{\,}_{1}$
and $\mathcal{P}^{\,}_{2}$ was given in
Fig.\ \ref{fig: Example lattice 2D model}.
Four more examples and one counter example
are given in Fig.~\ref{fig: notrivial examples path}.

\subsection{Relation to the construction for scar states from
Ref.~\cite{shiraishi2017systematic}}

In this section, we show that there exists a unitary transformation that
brings Hamiltonian~\eqref{eq:1a Hamiltonian} with the property
\eqref{eq:1b Hamiltonian}
into the form of the family of Hamiltonians defined
in Eqs.\ (1) and (2) from Ref.~\cite{shiraishi2017systematic}.
However, we emphasize that Hamiltonian~\eqref{eq:1a Hamiltonian}
stems from the stochastic matrix form Hamiltonians introduced in
Refs.~\cite{Castelnovo2005}, wherein
the property
\eqref{eq:1b Hamiltonian}
was proven.

We present the local Hermitian operator
$Q^{\,}_{s}$ in Eq.~\eqref{eq:1a Hamiltonian}
(the $\beta$ dependence is implicit) as
\begin{subequations}
\begin{equation}
Q^{\,}_{s} =
\sum_{a(s)}
\lambda^{\,}_{a(s)}\,
|\psi^{\,}_{a(s)}\rangle\langle\psi^{\,}_{a(s)}| \, ,
\end{equation}
where $a(s)$ labels the orthogonal eigenstates $|\psi^{\,}_{a(s)}\rangle$
with the real-valued eigenvalues $\lambda^{\,}_{a(s)}$ of $Q^{\,}_{s}$.
The consequence of the locality of $Q^{\,}_{s}$, in this paper, is that
its spectrum is bounded and discrete.
Moreover, by construction, $Q^{\,}_{s}$ has zero eigenvalues.
We denote by $\mathcal{T}(s)$ the kernel of $Q^{\,}_{s}$,
i.e., the subspace spanned by the eigenvectors with vanishing eigenvalues
$\lambda^{\,}_{a'(s)}=0$.
[From here, we use primed label $a'(s)$ for $a'(s) \in \mathcal{T}(s)$
and unprimed label $a(s)$ for $a(s) \notin \mathcal{T}(s)$.]
We shall define the local projector
\begin{equation}
P^{\,}_{s}\:=
\sum_{a(s) \, \notin \, \mathcal{T}(s)}
|\psi^{\,}_{a(s)}\rangle\langle\psi^{\,}_{a(s)}|
\label{eq: shiraishi projector}
\end{equation}
\end{subequations}
that assigns to all eigenspaces of $Q^{\,}_{s}$
with nonzero eigenvalue the eigenvalue 1.
The eigenvalue of the null state $|\Psi(\beta)\rangle$
with respect to both $P^{\,}_{s}$ and $Q^{\,}_{s}$ is $0$ for all $s$.  
We define the local Hermitian operator
\begin{subequations}
\begin{equation}
\widetilde{Q}^{\,}_{s}\:= 
\sum_{a(s) \, \notin \, \mathcal{T}(s)}
\lambda^{\,}_{a(s)}\,
|\psi^{\,}_{a(s)}\rangle\langle\psi^{\,}_{a(s)}| 
+
U
\sum_{a'(s) \, \in \, \mathcal{T}(s)}
|\psi^{\,}_{a'(s)}\rangle\langle\psi^{\,}_{a'(s)}| 
\end{equation}
together with the counterpart to Eq.~\eqref{eq:1a Hamiltonian} defined by
\begin{equation}
\widetilde{H}\:=
\sum_{s}
\alpha^{\,}_{s}\,
\widetilde{Q}^{\,}_{s}=
\sum_{s}
\sum_{a(s) \, \notin \, \mathcal{T}(s)}
\alpha^{\,}_{s}\,
\lambda^{\,}_{a(s)}\,
|\psi^{\,}_{a(s)}\rangle\langle\psi^{\,}_{a(s)}| 
+
U\,
\sum_{s'}
\sum_{a'(s') \, \in \, \mathcal{T}(s')}
\alpha^{\,}_{s'}\,
|\psi^{\,}_{a'(s')}\rangle\langle\psi^{\,}_{a'(s')}|\equiv
\sum_{s}
P^{\,}_{s}\,\tilde{h}^{\,}_{s}\,P^{\,}_{s}
+
\widetilde{H}',
\label{eq: shiraishi form}
\end{equation}
where
\begin{equation}
\tilde{h}^{\,}_{s}\:=
\alpha^{\,}_{s}
\sum_{a(s)}
\lambda^{\,}_{a(s)}\,
|\psi^{\,}_{a(s)}\rangle\langle\psi^{\,}_{a(s)}|,
\qquad
\widetilde{H}'\:=
U\,
\sum_{s'}
\sum_{a'(s') \, \in \, \mathcal{T}(s')}
\alpha^{\,}_{s'}\,
|\psi^{\,}_{a'(s')}\rangle\langle\psi^{\,}_{a'(s')}|.
\end{equation}
\end{subequations}

The projector defined by Eq.~\eqref{eq: shiraishi projector} and 
$\widetilde{H}'$ fulfill all the conditions of their counterparts
in Eqs.\ (1) and (2) from Ref.~\cite{shiraishi2017systematic},
respectively. Since $U\in\mathbb{R}$ is allowed to take the value $0$,
in which case $\widetilde{Q}^{\,}_{s}=Q^{\,}_{s}$, $\widetilde{H}=H$,
and $[\widetilde{H}',P^{\,}_{s}]=0$,
our Hamiltonian $H$ in Eq.~\eqref{eq:1a Hamiltonian} belongs to the
family of Hamiltonians defined by Ref.~\cite{shiraishi2017systematic}.

\end{widetext}


\end{document}